\documentclass[twocolumn,showpacs,preprintnumbers,amsmath,amssymb]{revtex4}

\usepackage{graphicx}
\usepackage{dcolumn}
\usepackage{bm}
\usepackage{natbib}

\begin{document}


\title{Quantum Pair Creation of Soliton Domain Walls}

\author{J. H. Miller, Jr.}
 \email{jhmiller@uh.edu}
\author{G. C\'{a}rdenas}
\author{A. \surname{Garc\'{\i}a P\'{e}rez}}
\author{W. More}
\author{A. W. Beckwith}
\affiliation{\ \\Department of Physics and Texas Center for
Superconductivity and Advanced Materials\\ University of Houston\\
Houston, Texas 77204-5005, USA}

\date{\today}

\begin{abstract}
A large body of experimental evidence suggests that the decay of
the false vacuum, accompanied by quantum pair creation of soliton
domain walls, can occur in a variety of condensed matter systems.
Examples include nucleation of charge soliton pairs in density
waves [eg. J. H. Miller, Jr. et al., Phys. Rev. Lett. {\bf 84},
1555 (2000)] and flux soliton pairs in long Josephon junctions.
Recently, Dias and Lemos [J. Math. Phys. {\bf 42}, 3292 (2001)]
have argued that the mass $m$ of the soliton should be interpreted
as a line density and a surface density, respectively, for (2+1)-D
and (3+1)-D systems in the expression for the pair production
rate. As the transverse dimensions are increased and the total
mass (energy) becomes large, thermal activation becomes
suppressed, so quantum processes can dominate even at relatively
high temperatures. This paper will discuss both experimental
evidence and theoretical arguments for the existence of
high-temperature collective quantum phenomena.
\end{abstract}


\maketitle

\section{ Background and Motivation}
A wide class of nonperturbative phenomena in field theory can be
understood in terms of quantum tunneling. A well-known example is
the quantum decay of the false vacuum \cite{Coleman}, which has
been of broad scientific interest in cosmology \cite{Tryon,Linde}
and other fields \cite{Caldeira} for over two decades. In three
dimensions, the boundary between the bubble of true vacuum and the
surrounding false vacuum is a type of topological defect known as
a domain wall. A variety of topological defects in condensed
matter systems have been proposed to nucleate via quantum or
thermal fluctuations. These include vortex-antivortex pairs and
vortex rings in superconductors \cite{Kosterlitz,Duan},
superfluids \cite{Rayfield,Volovik}, and Bose-Einstein condensates
\cite{Anderson}; dislocation pairs in Wigner crystals \cite{Chui}
and vortex lattices \cite{Blatter}; phase-slip vortex rings in
charge density waves (CDWs) \cite{Zaitsev-Zotov,Matsukawa}; charge
(or flux) soliton-antisoliton pairs in density waves \cite{Maki,
Maiti} (or Josephson junctions \cite{Widom}); and soliton-like
domain walls surrounding cigar-shaped bubbles of true vacuum in
three-dimensional CDWs \cite{Krive1}.

Topological defects, such as flux vortices, play an especially
important role in the cuprates and other type-II superconductors.
Magnetic relaxation rates that depend weakly on temperature up to
20 K \cite{Wen}, or even decrease with temperature \cite{Xue},
suggest that vortices may tunnel over a wide temperature range.
Moreover, the consistently low $I_{c}R_{n}$ products of cuprate
Josephson devices suggest that Josephson vortex-antivortex pair
creation \cite{Widom} may occur when the current is much smaller
than the ``classical" critical current $I_{0} \sim \Delta /R_{n}
e$.

A charge density wave (CDW) is a condensate \cite{Gruner} in which
the electronic charge density in a quasi-one-dimensional metal is
modulated,
$\rho(x,t)=\rho_{0}(x,t)+\rho_{1}\cos[2k_{F}x-\phi(x,t)]$. Here
$\rho_{0}(x,t)$ contains the background charge of the condensed
electrons, and an excess or deficiency of charge proportional to
$\pm \partial \phi/\partial x$. A spin density wave (SDW) has a
modulated spin density, and is equivalent to two out-of-phase CDWs
for the spin-up and spin-down subbands. Although pinned by
impurities, a density wave (DW) can transport a current when an
applied field exceeds a threshold value $E_{T}$. The most widely
studied models of DW depinning are variations of that proposed by
Fukuyama, Lee, and Rice \cite{Fukuyama}.

A long-standing debate concerns whether a classical description is
sufficient to describe density wave depinning, or a quantum
treatment is required. John Bardeen \cite{Bardeen1,Bardeen2}
proposed a model in which condensed, dressed electrons Zener
tunnel through a ``pinning gap," whose energy per electron is
small compared to the Peierls gap. This model was motivated by the
Zener-like behavior, $J \sim [E-E_{T}]\exp(-E_{0}/E)$
\cite{Monceau,Thorne1} observed in the current-field
characteristics of NbSe$_{3}$ and TaS$_{3}$. Bardeen pointed out
that the 3-D coherence of the DW condensate would suppress thermal
excitations without necessarily suppressing the amplitude for
coherent tunneling of the internal microscopic degrees of freedom.
The observed Zener field $E_{0}$ does not increase with the
transverse dimensions, which has important implications for any
theory based on tunneling probabilities.

Maki \cite{Maki} considered a real-space picture, in which
$2\pi$-soliton-antisoliton ($S-S'$) pairs nucleate via quantum
tunneling. Krive and Rozhavsky \cite{Krive2} pointed out the
existence of a Coulomb blockade threshold field for the creation
of charged $S-S'$ pairs, and later extended this picture to
nucleation of soliton-like domain walls \cite{Krive1}. A more
recent paper \cite{Miller1} proposed an analogy to time-correlated
single electron tunneling to explain the observed lack of DW
polarization below threshold, and to interpret key features of DW
dynamics, such as coherent oscillations and narrow-band noise.
Observations of Aharonov-Bohm oscillations \cite{Latyshev} in the
magneto-conductance of CD Ws in NbSe$_{3}$ crystals with columnar
defects strongly support the idea that quantum transport plays a
fundamental role. Additional evidence for quantum behavior in
density waves is found in \textit{rf\/} experiments
\cite{Miller2,Miller3} that show very good agreement with
photon-assisted tunneling theory at temperatures up to 215 K.

It is important to also consider whether or not density wave
dynamics can be interpreted classically. Perhaps the most obvious
requirement for classical sliding is that the washboard potential
should be tilted enough for the deformable object to slide over
the pinning barrier. This tilting of the washboard potential leads
to a predicted phase displacement of $\pi/2$ in the s-G model at
the classical threshold, and to even greater predicted phase
displacements when disorder is included \cite{Littlewood}.
However, NMR and high-resolution x-ray scattering experiments show
that the phase displacements of both charge
\cite{Ross1,Ross2,Requardt} and spin \cite{Wong} density waves are
more than an order of magnitude smaller than classically predicted
values near threshold.

Classical models also predict \cite{Littlewood,Narayan} that the
low-frequency dielectric response should increase, or even
diverge, as the classical threshold is approached from below. Such
predictions are refuted by the observed bias-independent
\textit{rf\/} and microwave responses below threshold in both CDWs
\cite{Miller3,Zettl} and Wigner crystals \cite{Li}. These
experiments strongly suggest that the washboard pinning potential
is barely tilted as the measured threshold field is attained.
Thus, the observed threshold appears to be much smaller than the
classical depinning field, and may represent a Coulomb blockade
threshold for the creation of charged topological defects.
Moreover, attempts \cite{Gammie} using a scanning tunneling
microscope (STM) to directly observe either displacement of the
CDW below threshold or sliding above threshold have been
unsuccessful. The apparent lack of sliding seen in STM experiments
and the jerky dynamics revealed by NMR experiments
\cite{Ross1,Ross2} suggest that the DW spends most of its time in
the pinned state even above threshold.

The classical deformable DW model also predicts that the
high-field \textit{dc\/} conductivity should have the form
$\sigma_{cdw} \sim a-bE^{-1/2} $, where $a$ and $b$ are constants
\cite{Sneddon}. This prediction is contradicted by experiments
\cite{Thorne2} that show substantial departure from the
classically predicted behavior, but essentially perfect agreement
with the Zener form $\sim \exp (-E_{0}/E)$, in the high-field
limit. Coppersmith and Littlewood \cite{Coppersmith} (CL) make the
following predictions for mode locking of a moving DW to an
\textit{ac\/} source within the classical model. (1) Damped
relaxation is crucial in that complete mode locking can occur only
when the \textit{ac\/} frequency and amplitude are small enough
for significant relaxation to take place when the total field is
less than $E_{T}$. (2) For large \textit{ac\/} amplitudes and
frequencies such that the time interval spent below threshold is
small, the differential resistance $dV/dI$ should exhibit peaks
but not complete mode locking. In addition, no ``wings" (sharp
negative dips in $dV/dI$ adjacent to the interference peak) should
be observed. These predictions are refuted incontrovertibly by
\textit{ac-dc\/} interference and mode locking experiments on high
quality NbSe$_{3}$ crystals \cite{Thorne2} that demonstrate both
negative wings and complete mode locking at high \textit{dc\/}
bias fields and high frequencies.

The failure of CL's calculations to account for the mode-locking
experiments reflects a fundamental deficiency of the classical
deformable model. When the DW slides rapidly, relaxation into the
pinning potential wells cannot occur quickly enough to prevent the
deformations and net pinning energy from vanishing. The
characteristic frequency for this relaxation, the DW dielectric
relaxation frequency, is measured to be about 5 MHz in NbSe$_{3}$
\cite{Miller4}. The observations of complete mode locking at
\textit{ac\/} and drift frequencies of up to 1 GHz indicate that
the strength of DW pinning is undiminished in a regime where
significant relaxation cannot occur within the conventional
classical model. Thus, quantum mechanisms must be explored to
understand, fundamentally, DW depinning and dynamics.

\section{Soliton Tunneling Model}
A realistic density wave pinning potential would have the form
$U_{0}(\mathbf{r})\{1-\cos[\phi-\phi_{0}(\mathbf{r})]\}$, which
includes spatial variations of both the pinning energy $U_{0}$,
and the optimum phase $\phi_{0}$. However, observations of narrow
band noise, coherent oscillations, and complete mode-locking with
an \textit{ac\/} source \cite{Thorne2} in high quality NbSe$_{3}$
crystals suggest that both $U_{0}$ and $\phi_{0}$ are slowly
varying in weakly pinned DWs. The sine-Gordon (s-G) model thus
provides an idealized conceptual framework for interpreting many
aspects of density wave transport \cite{Rice}, from both classical
and quantum points of view. Similarly, a weakly-coupled Josephson
junction (JJ) can be described by a phase, $\phi(x,t)$,
representing the phase difference across the junction, in a
sine-Gordon potential.

\begin{table*}[t!]
  \caption{Charge-flux duality between density waves and Josephson
  junctions.}
  \label{Tab:Charge-flux-duality}
  \begin{ruledtabular}
  \begin{tabular}{lcc}
   & Density Wave & Josephson junction \\
   &  &  \\
  Soliton or antisoliton & Kink w/ charge $\pm Q_{0}$ & Josephson vortex w/ flux $\pm \phi_{0}$ \\
   &  &  \\
  Type of threshold & Threshold field $E_{T}$ & Threshold current $I_{T}$ \\
   &  &  \\
  Transport characteristic & $I\; vs.\; V$, $I = 2\pi Q_{0} \partial \phi/\partial t$ & $V\; vs.\; I$, $V = 2\pi \phi_{0} \partial \phi/\partial t$ \\
  \end{tabular}
  \end{ruledtabular}
\end{table*}

Density waves and JJs are dual in that the roles of charge and
flux are interchanged, as well as those of current and voltage.
The current in a density wave is $I=(Q_{0}/2\pi)\partial\phi
/\partial t$, where $Q_{0} \sim 2eN_{ch}$ and $N_{ch}$ is the
number of parallel chains, whereas the voltage across a JJ is
$V=(\Phi_{0}/2\pi)\partial\phi /\partial t$, where
$\Phi_{0}=h/2e$. Charge (flux) solitons in a density wave (JJ)
carry a charge (flux) of $\pm Q_{0} (\pm \Phi_{0})$ (see Table
\ref{Tab:Charge-flux-duality}). The width of a Josephson vortex is
roughly the Josephson penetration length, $\lambda_{J}\propto
J_{c}^{-1/2}$ whereas, in a DW, the soliton width is
$\lambda_{0}=c_{0}/\omega_{0}$, where $c_{0}$ is the phason
velocity and $\omega_{0}$ is the pinning frequency. Thus,
$\lambda_{0}$ will increase with decreasing impurity concentration
(as $\omega_{0}$ decreases), and may approach the distance between
contacts in extremely pure samples. This is equivalent to
approaching the short junction limit, $L < \lambda_{J}$, in a JJ,
where the $V-I$ curves become significantly less rounded.

The quantum decay of the metastable false vacuum of a scalar field
$\phi$, accompanied by the creation of solitons and antisolitons
in (1+1) dimensions \cite{Stone,Voloshin,Kiselev} has been studied
extensively in quantum field theory. Dias and Lemos \cite{Dias1}
(DL) have calculated the effective one loop action for pair
creation of solitons in the (1+1)-D s-G model. In addition, they
have extended their arguments to pair creation of soliton domain
walls in higher dimensions.  In the (1+1)-D case, the
soliton-antisoliton pair production rate per unit time and length,
$\Gamma/L$, is identified with the decay rate of the false vacuum.
This is calculated to be, in the absence of Coulomb interactions
between the solitons (taking $\hbar=c=1$) \cite{Dias1}:

\begin{equation}\label{Eq:Pair-prod-rate}
  \frac{\Gamma}{L}=\frac{\varepsilon}{2\pi} \exp\left[\frac{-\pi m^{2}}{\varepsilon}\right]
\end{equation}

\noindent where $m$ is the soliton mass (energy). Here
$\varepsilon$ represents the gain in energy per unit length
between the two created solitons, which is proportional to the
applied electric field $E$ for a density wave (and to the applied
current for a JJ). The energy for pair creation is contained in
the line between the particles and is given by $2x\varepsilon
=2m$, where $2x$ is the distance between the solitons.

DL point out \cite{Dias1} that a one-particle system in (1+1)-D
can be transformed into a line in (2+1)-D and a thin wall in
(3+1)-D, except that the mass $m$ should be interpreted as a line
density and a surface density, respectively. Their calculations
thus apply directly to the domain wall pair creation. As the
transverse dimensions are increased and the total mass (energy)
becomes large, thermal activation becomes suppressed, so quantum
processes can dominate even at relatively high temperatures (up to
400 K for the CDW in NbSe$_{3}$).

As an example, consider the (2+1)-D case, where one can make a cut
normal to the infinite strings so that ``$\varepsilon$" will still
be the energy density of the line between the two infinitesimal
pieces of the strings. Each one of these infinitesimal pieces has
a mass $m$, so $m$ represents the line density of the string. The
energy for pair creation of strings comes from the infinite plane
between the two strings, and is given by $2xd\varepsilon = d2m$,
where $d$ is the length of the string, which goes to infinity
\cite{Dias2}. Therefore both the mass of the string and the energy
to create the pair of strings become infinite.

Importantly, ``$m$" can no longer be interpreted as the total mass
of the string or of the domain wall. Indeed such an interpretation
would correspond to creation of point particles in (2+1)-D or in
(3+1)-D, where the total mass $m$ would be infinite. The infinite
energy for pair creation would then have to come from the line
between the two particles rather than from the infinite plane
between the two strings. DL's calculations do not apply to this
process. In the generalization to higher dimensions, one is using
the symmetry of the system along the extra direction. In Minkowski
space-time, the procedure is direct \cite{Dias2}. However, in
systems that include gravitation, the procedure is usually not so
straightforward, and one must employ dimensional reduction or
dimensional expansion \cite{Lemos}.

Suppose we now put the constants $\hbar$ and $c$ back into Eq.
\ref{Eq:Pair-prod-rate} and take $\lambda$ to represent the
Compton wavelength, $\lambda=\hbar/mc$. (In a density wave, the
effective ``speed of light" is the phason velocity $c_{0}$, which
is about $3\times 10^{5}$ cm/s.) Explicitly putting the physical
constants into the exponent, we then have: $m\pi^{2}/\varepsilon
\rightarrow m\pi c^{2}/[qE\lambda]$, where $q$ is the soliton
charge for a single chain (i.e. for the (1+1)-D system) and $E$ is
the electric field. (In a JJ, the charge and electric field would
be replaced by the flux $\Phi_{0}$ and the current density.) For
the (3+1)-D system, we can now define $m'=m/a^{2}$ and
$q'=q/a^{2}$ to be the mass and charge per unit area of the
soliton domain wall, respectively. In a density wave, the length
represents the transverse distance between DW chains. Note that,
for the exponent $\pi m'c^{2}/[q'E\lambda]$ to remain
dimensionless, $\lambda$ needs to retain its original units of
length. One way of doing this is to define a new ``quantum of
action per unit area" in terms of the characteristic length scale
$a$, i.e. $\hbar'=\hbar/a^{2}$ , leading to a Compton wavelength,
$\lambda=\hbar'/m'c$, which has the same value as before. Lattice
gauge theorists have recently examined models of domain wall
fermions (bosons), whose anticommutators (commutators) differ from
those of point particles, thus lending credence to the arguments
presented here.

Another way of looking at the problem is to recall that $\hbar$
has units of angular momentum: $L = rmv$, where $r$ = distance,
$m$ = mass, and $v$ = velocity. Now, since the mass transforms as
$m' = m/A$, where $A$ is an area, from $L = rmv$ we must have $L'
= L/A$, so that $\hbar'=\hbar/A$ \cite{Dias2}. The commutator
between bosonic domain wall creation and annihilation operators
will then be scaled accordingly. The Compton wavelength of the
(3+1)-D problem refers to the direction normal to the domain wall,
and is thus $\lambda=\hbar'/m'c=\hbar/mc$.

It should be noted that \textit{rf\/} experiments
\cite{Thorne1,Miller2,Miller3} are consistent with a ratio
$\hbar/q=\hbar'/q'$ that is scale invariant, where $q = 2e$ for a
fully condensed CDW. This is further supported by
magneto-transport experiments \cite{Latyshev}, which yield
Aharonov-Bohm oscillations with a periodicity of $h/2e$, and not
$h/2Ne$ as predicted theoretically \cite{Bogachek}, where $N$ is
the number of coupled chains. However, the observed mode locking
with an \textit{ac\/} source \cite{Thorne2} shows that the phase
is coherent throughout macroscopic regions within the crystal,
suggesting quantum nucleation of entire domain walls rather than
just single dislocations.

The above arguments lead, for a CDW, to a Zener-like current-field
characteristic given by:

\begin{equation}\label{Eq:Zener-like-current}
  J \sim E \exp[-E_{0}/E]
\end{equation}

\noindent where the characteristic Zener field $E_{0}=[\pi
m'^{2}c_{0}^{3}]/[\hbar'q']$ is independent of the traverse
dimensions. (For a JJ, the $V\; vs.\; I$ curve would be given by
$V \sim I\exp[-I_{0}/I]$.) DL's calculations neglect Coulomb
interactions between charge solitons (or magnetic interactions
between flux solitons in a JJ), and thus fail to predict a sharp
threshold field (threshold current for a JJ) for pair creation.
The following section examines the origin of the threshold field
as a macroscopic Coulomb blockade effect (or its dual for the case
of a JJ).

\section{Macroscopic Coulomb Blockade and Time-Correlated Soliton
Tunneling} 
\begin{figure}[t!]
  \includegraphics[scale=0.3]{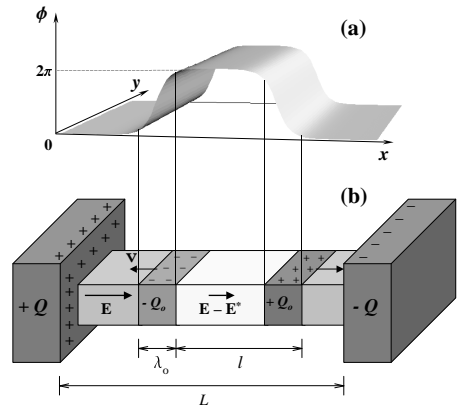}
  \caption{(a) Density wave phase vs. position, illustrating the
  production of a soliton-antisoliton domain wall pair. (b) Model
  of a density wave capacitor, showing the nucleated domain walls
  moving towards the contacts. The electric field between the domain
  walls is reduced by the internal field $E^{*}$. The distances $l$,
  $\lambda_{0}$, and the crystal thickness are greatly exaggerated
  for clarity.}
  \label{Fig:Phase-position}
\end{figure}

The quantum interpretation of the threshold field, as a
pair-creation threshold due to Coulomb blockade, is motivated by
Coleman's paper \cite{Coleman2} on soliton pair-creation in the
massive Schwinger model. A pair of $S$ and $S'$ domain walls with
charges $\pm Q_{0}$ produce an internal field of magnitude $E^{*}
= Q_{0}/\varepsilon A$, as shown in Fig. \ref{Fig:Phase-position},
where $A$ is the cross-sectional area and $\varepsilon$ is the
dielectric constant. When a field $E$ is applied, the difference
in electrostatic energies of a state with a pair of separation $l$
and of the ``vacuum" is $\Delta U=(1/2)\varepsilon A l
[(E+E^{*})^{2}-E^{2}]=Q_{0}l[(1/2)E^{*} \pm E]$ , which is
positive when $|E|<(1/2)E^{*}$. Conservation of energy thus
forbids pair production for fields less than a quantum threshold,
$E_{T}\equiv (1/2)E^{*}=Q_{0}/2\varepsilon A$, which appears to be
about two orders of magnitude smaller than the classical depinning
field in NbSe$_{3}$ \cite{Miller1}. The observed universality
relation, $\varepsilon E_{T} \sim eN_{ch}/A$ \cite{Gruner}, thus
arises quite elegantly in this model. Screening by the uncondensed
normal carriers greatly enhances the dielectric response of
NbSe$_{3}$, which has an incomplete Peierls gap and also has a
much lower threshold field than most other CDW materials. (In a
JJ, charge-flux duality leads to a quantum threshold current
proportional to $\Phi_{0}/2L_{M}$, where $L_{M}$ is the mutual
inductive coupling between the Josephson vortex and the
antivortex.)

\begin{figure}[t!]
\vspace*{-0.5cm}
  \setlength{\abovecaptionskip}{-30pt}
  \includegraphics[scale=0.3]{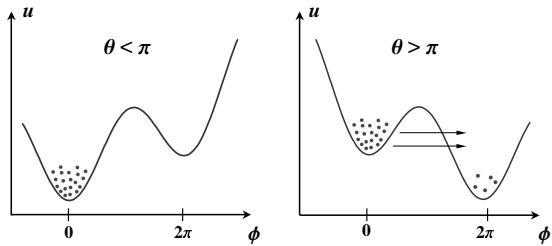}
  \caption{Plot of potential energy vs. $\phi$ for two different values
  of $\theta$, in which the many spatial degrees of freedom are illustrated
  schematically. Tunneling is prevented by conservation of energy
  when $\theta < \pi$. When $\theta > \pi$, parts of the system tunnel
  into the adjacent well via decay of the false vacuum.}
  \label{Fig:Energy-phi}
\end{figure}

A density wave between two contacts behaves as a capacitor with an
enormous dielectric constant, as shown in Fig.
\ref{Fig:Phase-position}. The initial charging energy is
$Q^{2}/2C$, where $Q$ is the displacement charge and $C =
\varepsilon A/L$. We define $\theta \equiv 2\pi Q/Q_{0}=2\pi
E/E^{*}=\pi E/E_{T}$ and note that a displacement $\phi$ near the
middle creates a non-topological kink-antikink pair, with charges
$\pm (\phi/2\pi)Q_{0}$, if $\phi=0$ at the contacts. The washboard
pinning and quadratic charging energies can then be written as
\cite{Miller1}:

\begin{equation}\label{Eq:Washboard-pinning}
  U[\phi]=\int_{0}^{L}dx \;\{u_{p}[1-\cos \phi(x)]+u_{c}[\theta - \phi(x)]^{2}\}
\end{equation}

\noindent where the first and last terms represent the pinning and
electrostatic charging energies, respectively, and where $u_{p}
\gg u_{c}$ for NbSe$_{3}$. If the system starts out in its ground
state, conservation of energy will prevent tunneling when the
applied field is below threshold, $\theta < \pi \;(E < E_{T})$, as
illustrated in Fig. \ref{Fig:Energy-phi}. However, when $\theta$
exceeds $\pi$, what was formerly the true vacuum becomes the
unstable false vacuum. One or more bubbles of true vacuum, with
soliton domain walls at their surfaces, then nucleate and expand
rapidly (Fig. \ref{Fig:Phase-position}). After $n$ solitons of
charge $Q_{0}$ (and antisolitons of charge $-Q_{0}$) have reached
the contacts, the charging energy becomes:

\begin{equation}\label{Eq:Charge-energy}
\frac{(Q-nQ_{0})^2}{2C}=\frac{{Q_{0}}^2}{8\pi^2 C}(\theta-2\pi
n)^2
\end{equation}

This series of piecewise parabolas is similar to the charging
energy of a single-electron tunnel junction, except that $Q_{0}$
now represents a macroscopic charge.

The single-electron transistor (SET) \cite{Devoret} consists of a
gate capacitor $C_{g}$ coupled to an island electrode between two
small capacitance tunnel junctions in series. The gate voltage
modulates the $I-V$ curves between the source and drain
electrodes, with a period $e$ in displacement charge, $Q_{g} =
C_{g}V_{g}$. The displacement charges $Q_{1,2}$ across the two
tunnel junctions are related as $Q_{2} = Q_{1} + Q_{g} + q_{0}$,
where $q_{0}$ is a phenomenological offset charge induced during
cooling \cite{Devoret}. The SET is related by charge-flux duality
to the dc SQUID. The critical voltage across an SET is a periodic
function of $Q_{g}$, whereas the critical current across a SQUID
is periodically modulated (with period $\Phi_{0}$) by the flux
$\Phi$.

The model discussed above suggests that it may be possible to
demonstrate a macroscopic version of the SET, by attaching a gate
capacitor to an island electrode near the center of a quasi-1-D
crystal with a density wave. The displacement charge induced by
the gate electrode would then periodically modulate the total
critical voltage between the source and drain electrodes. Ideally,
in the absence of any shunt conductance, the periodicity of the
gate displacement charge might be expected to be $\sim Q_{0}$.
However, screening by the normal, uncondensed electrons will tend
to reduce the effectiveness of the gate which, unlike the source
and drain contacts, cannot be driven by a current source. The
displacement charges across the two segments of the crystal will
be related as $Q_{2}=Q_{1}+\beta (Q_{g}+q_{0})$, where $\beta \ll
1$ reflects screening by the normal carriers. The total charging
energy of the two segments, in this idealized model, will then be:
\begin{widetext}
\begin{equation}\label{Eq:Total-charge-energy}
  \frac{(Q_{1}-n_{1}Q_{0})^2}{2C_{1}}+\frac{(Q_{2}-n_{2}Q_{0})^2}{2C_{2}}=\frac{{Q_{0}}^2}{8\pi^2
  C}\left\{\frac{(\theta_{1}-2\pi n_{1})^2}{2C_{1}} +\frac{(\theta_{2}-2\pi n_{2})^2}{2C_{2}}\right\}
\end{equation}
\end{widetext}
\noindent where $C_{1}$ and $C_{2}$ are the capacitances of the
two segments separated by the island electrode. The analogy to the
SET suggests that a gate voltage might modulate the $I-V$ curves
between source and drain contacts, with a periodicity $\Delta
V_{g} \sim Q_{0}/\beta C_{g}$. The gate capacitance $C_{g}$, and
hence the attainable displacement charge $Q_{g}$, may have been
too small to observe non-monotonic behavior in previous
experiments \cite{Adelman}, in which a gate electrode was
fabricated directly on the crystal to form a MOSFET-like
structure. The soliton tunneling transistor (STT), discussed in
the next section, employs a much larger, $1\; \mu$F, gate
capacitor coupled to an NbSe$_{3}$ crystal, and exhibits
non-monotonic behavior.

An exact calculation of the charging energy would yield a plot in
which the energy is reduced slightly at the crossing points,
$\theta = n\pi$, as compared to piecewise parabolas. However, Eq.
\ref{Eq:Charge-energy} provides a reasonable approximation to
$E(\theta)$:

\begin{equation}\label{Eq:E-approx}
  E(\theta) \sim (\theta - 2\pi n)^2
\end{equation}

As in the SET, the voltage across each segment of the STT is
related to the charging energy as $V_{1,2}=dE/dQ_{1,2} \propto
dE/d\theta_{1,2}$. If we use the approximation given by Eq.
\ref{Eq:E-approx}, this yields a sawtooth function, which can be
expanded as a Fourier series:

\begin{equation}\label{Eq:Sawtooth}
  V_{1,2}=V_{0}\,{\rm saw}(\theta_{1,2})=-\frac{V_{0}}{\pi}
  \sum_{n=1}^{\infty}\frac{(-1)^n}{n}\sin(n\theta_{1,2})
\end{equation}

This model can also been used to model the dynamics of density
waves and to interpret the narrow-band noise spectra, the many
harmonics of which are consistent with a sawtooth function.

\begin{figure}[t!]
  \includegraphics[scale=0.3]{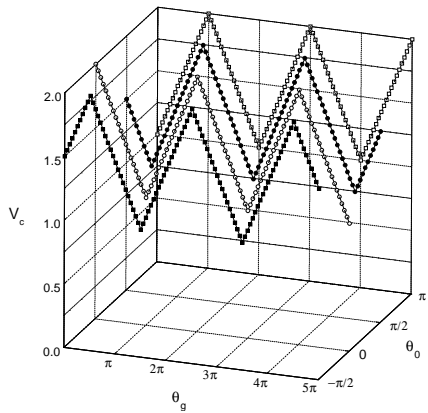}
  \caption{Predicted critical voltage vs. normalized gate voltage
  $\theta_{g}$ for several values of $\theta_{0}$ using the idealized
  model discussed in Section III, showing the periodic behavior.}
  \label{Fig:Vcrit-Vg}
\end{figure}

Equation \ref{Eq:Sawtooth} is related by charge-flux duality to
the current-phase relation of a Josephson junction. The periodic
behavior of the source-to-drain voltage vs. gate voltage of an STT
can readily be understood by exploiting the duality with a dc
SQUID. When each JJ in a dc SQUID has an ideal, sinusoidal
current-phase relation and the critical currents $I_{0}$ are
identical, the total current is
$I=I_{0}[\sin\varphi_{1}+\sin\varphi_{2}]$, where
$\varphi_{2}=\varphi_{1}+2\Phi/\Phi_{0}$. This yields a total
critical current, $I_{c} = 2I_{0}|\cos(2\pi\Phi/\Phi_{0})|$, which
is a periodic function of the flux $\Phi$. Similarly, the total
source-to-drain voltage of an ideal STT, $V = V_{1} + V_{2}$,
where $V_{1,2}$ are given by Eq. \ref{Eq:Sawtooth}, yields a
critical voltage that is a periodic function of gate voltage
$V_{g} = Q_{g}/C_{g}$, as shown in Fig. \ref{Fig:Vcrit-Vg}. Here
we note that $\theta_{2} = \theta_{1} + \theta_{g} + \theta_{0}$,
where $\theta_{g} = 2\pi \beta Q_{g}/Q_{0}=2\pi\beta q_{0}/Q_{0}$,
$Q_{g}$ is the displacement charge across the gate capacitor,
$q_{0}$ is the offset charge, and $\beta$ is the screening
parameter due to the normal electrons.

\section{Soliton Tunneling Transistor Experiment}
The experiment reported here employs a geometry analogous to that
of the SET, in which the ``source-to-drain" $I-V$ characteristic
is modulated by a voltage applied to a gate capacitor. Single
crystals of NbSe$_{3}$ were employed in the experiment. This
material forms two independent CDWs, at Peierls transition
temperatures of 145 K and 59 K \cite{Monceau}, respectively. The
Peierls gap opens up over most of the Fermi surface (FS) below the
lower transition, but leaves a small portion of the FS intact, so
that a significant concentration ($\sim 6 \times 10^{-18}
cm^{-3}$) of normal, uncondensed carriers remain down to low
temperatures.

\begin{figure}[t!]
  \includegraphics[scale=0.3]{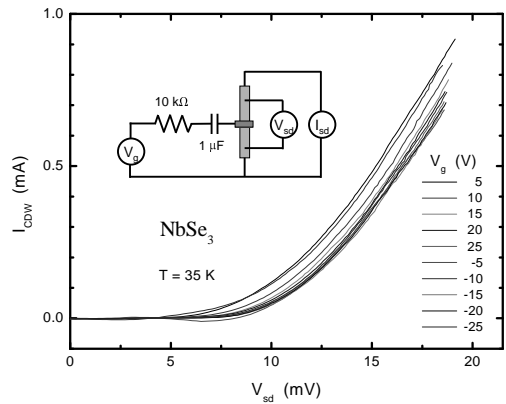}
  \caption{CDW current vs. source-to-drain voltage in a soliton tunneling
  transistor (NbSe$_{3}$) for several values of gate voltage at 35K. (The
  shunt current of the normal electrons has been subtracted for clarity.)}
  \label{Fig:Icdw-Vsd}
\end{figure}

The geometry used in our experiment is illustrated in the inset to
Fig. \ref{Fig:Icdw-Vsd}, where the width of the crystal is
exaggerated for clarity. The NbSe$_{3}$ crystal was placed onto an
alumina substrate with a series of evaporated, 25 $\mu$m wide gold
contacts. The substrate was thermally anchored to a cold-finger in
the vacuum shroud of an open cycle helium flow cryostat and the
temperature was controlled using a Lake Shore temperature
controller attached to a heater coil wrapped around the
cold-finger. A Keithley programmable \textit{dc\/} current source
injected the current between two contacts, which were bonded to
the crystal near the ends using silver paint. The
``source-to-drain" voltage was measured between two additional
gold contacts, as illustrated, and a 1 $\mu$F gate capacitor was
attached to the center gold contact using silver paint. The
spacing between contacts along the crystal was 500 $\mu$m
center-to-center, and the gate capacitor was kept inside the
cryostat.

We found that substantially smaller gate capacitors (as well as
gate capacitors with longer leads) were unable to induce a
periodic modulation of the $I-V$ characteristic. Moreover,
\textit{dc\/} $I-V$ (rather than differential $dV/dI$)
measurements were necessary to avoid inducing a displacement
current through the gate capacitor. A programmable voltage source
was coupled to the gate capacitor via a 10-k$\Omega$ resistor,
which limited the current flowing through the crystal during
changes in gate voltage when the gate capacitor either partially
charged or discharged. The cryostat was kept inside an
electromagnetically shielded enclosure, and $V_{sd}$ was measured
with a nanovoltmeter.

The measurements were primarily carried out at 35 K. Previous
``field effect transistor" experiments \cite{Adelman} showed the
greatest modulation at around 30 K, where the threshold field is
near its minimum. We also observed the largest modulation, using
our geometry, at comparable temperatures. We attained the best
temperature stability (better than $\pm 0.01$ K) when the
temperature was set to 35 K, and thus chose this temperature for
most of the measurements reported here.

\begin{figure}[t!]
  \includegraphics[scale=0.8]{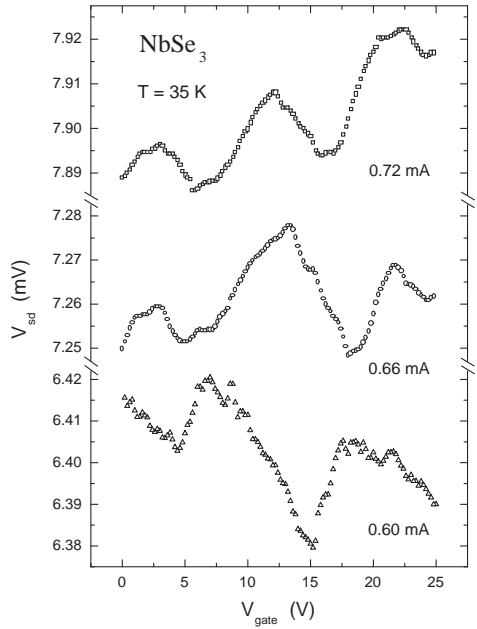}
  \caption{Source-to-drain voltage $V_{sd}\; vs.\;V_{g}$ for fixed values
  of total bias current at 35 K.}
  \label{Fig:Vsd-Vg}
\end{figure}

Figure \ref{Fig:Icdw-Vsd} shows several plots of CDW current as a
function of source-to-drain voltage, $I_{cdw}\; vs.\; V_{sd}$, in
a NbSe$_{3}$ crystal at 35 K, for different values of gate voltage
$V_{g}$. The gate voltage is seen to modulate the threshold
voltage in the $I-V$ curves of Fig. \ref{Fig:Icdw-Vsd}. Figure
\ref{Fig:Vsd-Vg} displays plots of source-to-drain voltage
$V_{sd}\; vs.\; V_{g}$ for three values of total bias current
above threshold. The plots exhibit roughly periodic behavior,
similar to that observed in SETs. However, in our system, the
measured periodicity $\Delta V_{g} \sim 10V$ is consistent with a
macroscopic displacement charge $\Delta Q = C_{g}\Delta V_{g} \sim
6 \times 10^{13} e$, comparable to the charge of the conducting
electrons between the contacts.

The behavior shown in Fig. \ref{Fig:Vsd-Vg} is quite
extraordinary, and appears to be consistent with the soliton
domain wall tunneling hypothesis. The number of parallel CDW
chains, $N_{ch}$, is about $10^{8}$. Thus, one might estimate the
screening parameter as follows: $\beta \sim Q_{0}/\Delta Q \sim
2N_{ch}e/\Delta Q \sim 3 \times 10^{-6}$. However, an alternative
interpretation for the observed periodicity might be that all of
the normal electrons between the contacts participate in screening
out the displacement charge $Q_{0}$ of the CDW. Thus, the
observation that $\Delta Q/e \sim 6 \times 10^{13}$ is roughly the
number of conducting electrons between contacts may not be a
coincidence. Further work is needed to better understand the
effects of screening by the normal, uncondensed electrons.

\section{Implications}
The implications of the model and experimental results reported
here are potentially far-reaching, and could impact fields as
diverse as cosmology, condensed matter physics, quantum
computation, and biophysics. In the field of cosmology, quantum
nucleation of dilaton black hole pairs \cite{Dowker}, cosmic
strings \cite{Basu}, and even the entire universe
\cite{Tryon,Linde} have been proposed. The European Science
Foundation has recently funded a program, known as Cosmology in
the Laboratory (COSLAB), intended to explore laboratory analogs of
cosmological objects. Topics being studied include analogies
between vortices in superconductors and cosmic strings, and
artificial (acoustic or optical) black holes.

In the area of biophysics, Fr\"{o}hlich suggested
\cite{Frohlich1,Frohlich2} that long-range quantum coherence may
play a fundamental role in biological systems. This may provide a
mechanism to integrate complex interactions that take place within
a live cell. A recent model \cite{Mavromatos} proposes that
microtubules, which exist in the cells of all higher organisms but
are especially concentrated in neurons, engage in a form of
quantum computation. Roger Penrose \cite{Penrose} has even made
compelling arguments that large-scale quantum coherence, perhaps
mediated by the microtubules, plays a fundamental role in
consciousness. It is clear that, for Penrose's arguments to be
valid, some mechanism is needed to suppress decoherence. Perhaps
soliton domain walls, or related topological defects, may prove to
be 'topologically protected' from decoherence at biological
temperatures. Further research is clearly warranted to explore
this extraordinary possibility.

\begin{acknowledgments}
The authors gratefully acknowledge the assistance of James
Claycomb, L.-M.(Patrick) Xie, and John McCarten as well as
valuable insights provided by Oscar Dias and Jose Lemos. This work
was supported, in part, by the State of Texas through the Texas
Center for Superconductivity and Advanced Materials and the Texas
Higher Education Coordinating Board Advanced Research Program
(ARP), and by the R. A. Welch Foundation (E-1221).
\end{acknowledgments}

\end{document}